\documentclass[preprint,showpacs,preprintnumbers,amsmath,amssymb]{revtex4}

\usepackage{graphicx}
\usepackage{dcolumn}
\usepackage{bm}
\usepackage{color}
\usepackage{float}

\usepackage{feynmf}
\usepackage{slashed}
\usepackage{enumerate}

\usepackage[utf8]{inputenc}

\begin{document}

\title{Lepton-portal Dark Matter in Hidden Valley Model and the DAMPE recent results}


\author{Yi-Lei Tang$^2$}

\author{Lei Wu$^1$}

\author{Mengchao Zhang$^3$}
\thanks{mczhang@ibs.re.kr}

\author{Rui Zheng$^4$}

\vspace{0.5cm}

\address{$^1$Department of Physics and Institute of Theoretical Physics, Nanjing Normal University, Nanjing, Jiangsu, 210023, China \\ $^2$School of Physics, KIAS, 85 Hoegiro, Seoul 02455, Republic of Korea \\ $^3$Center for Theoretical Physics of the Universe, Institute for Basic Science (IBS), Daejeon, 34126, Korea \\ $^4$Department of Physics, University of California, Davis, CA 95616, USA}

\date{\today}

\preprint{
CTPU-17-40
}

\begin{abstract}
We study the recent $e^\pm$ cosmic ray excess reported by DAMPE in a Hidden Valley Model with lepton-portal dark matter. We find the electron-portal can account for the excess well and satisfy the DM relic density and direct detection bounds, while electron+muon/electron+muon+tau-portal suffers from strong constraints from lepton flavor violating observables, such as $\mu \to 3 e$. We also discuss possible collider signatures of our model, both at the LHC and a future 100 TeV hadron collider.

\vspace{0.5cm}

\noindent\textbf{Keywords:} dark matter, DAMPE, LHC

\end{abstract}
\pacs{95.35.+d, 12.60.-i, 11.30.Hv}

\maketitle
\section{Introduction}

Gravitational effects have inferred the existence of dark matter (DM) in our universe. The weakly interacting massive particles (WIMPs), one of the most popular candidates of particle dark matter, have been investigated for decades. A wealth of experiments have been carried out in search of these elusive particles, but to no avail. Direct detection experiments aim to observe the scattering between the dark matter and the  nucleus. The recent strong limits have excluded spin-dependent cross sections above $\sim 2$-$4 \times 10^{-41} \text{cm}^2$ at the 90\% C.L. for a WIMP mass of $\sim 40 \text{GeV}$ \cite{PandaX_SD}, and spin-independent cross sections above $7.7 \times 10^{-47} \text{cm}^2$ for a WIMP mass of $35 \text{ GeV}$ \cite{xenon-1t}. Terrestrial experiments  searching for the mono-$X$ signatures at colliders also impose stringent bounds on various WIMP DM models. Besides, indirect detection via observing cosmic-rays, gamma-rays and neutrinos also provides a promising way to probe DM, such as PAMELA \cite{PAMELA}, AMS-02 \cite{AMS_02}, Fermi-LAT~\cite{Fermi_LAT} and CALET~\cite{CALET}.

Recently, the DArk Matter Particle Explorer (DAMPE) satellite experiment~\cite{dampemission} released their first results on the total flux of $e^{\pm}$ cosmic ray up to 5 TeV~\cite{dampe2017}. A tentative sharp peak at $\sim 1.4$ TeV is reported, but no related excess in the anti-proton flux is observed, which implies a nearby monoenergetic electron source. By fitting the energy spectrum of $e^{\pm}$, it is found that a sub-halo composed of 1.5 TeV DM $\rm 0.1 \sim 0.3$ kpc away from the solar system can account for such an $e^{\pm}$ cosmic ray excess~\cite{QiangYuan}. The corresponding DM annihilation cross section $<\sigma v>$ is around $3 \times 10^{-26} cm^3/s$. Several possible DM models for explaining DAMPE observation are proposed in~\cite{Zu:2017dzm,Gu:2017gle,Duan:2017pkq,Fang:2017tvj,Fan:2017sor,Athron:2017drj,Chao:2017emq,Huang:2017egk}.

In this paper, we interpret this DAMPE excess with a lepton-portal Dirac fermion DM in Hidden Valley Model. The spontaneous breaking of the dark gauge group leaves us with an unbroken global symmetry, which forbids the majorana mass terms and preserves the stability of the DM as well. The heavy scalar mediator that is charged under both the SM and dark gauge group connects the DM and SM leptons. We consider the constraints of DM relic abundance, direct detection and low energy observables, and we explore the potential of probing such lepton-portal DM through the process $pp \to 2\ell + E^{miss}_T$ at a 100 TeV hadron collider.  The structure of this paper is organized as follows. In Section \ref{section2}, we introduce our model. In Section \ref{section3}, we present numerical results on the various constraints and signals, and offer discussions. Finally, we draw our conclusions in Section \ref{section4}.

\section{Model}\label{section2}

In this paper, we consider a lepton-portal Dirac fermion DM in Hidden Valley Model with an extra $U(1)^{\prime}$ gauge symmetry. The corresponding effective Lagrangian is given by
\begin{eqnarray}
\mathcal{L} &=& \overline{\chi} ( i D\!\!\!\!/ - m_{\chi}) \chi - \frac{1}{4} F^{\prime \mu \nu} F^{\prime}_{\mu \nu} - \frac{1}{2} m_{A^{\prime}}^2 A^{\prime \mu} A^{\prime}_{\mu} \nonumber \\
&+& (D_{\mu} \phi) (D^{\mu} \phi)^{\dagger} - m_{\phi}^2 \phi \phi^* + \kappa_1 \phi \phi^* H^{\dag} H + \frac{\kappa_2}{2} \phi \phi^*\phi \phi^* \nonumber \\
&+& \lambda_i \phi \overline{\chi} P_R e_R^i + \text{h.c.}. \label{Lag}
\end{eqnarray}
Here, $\chi$ is the vector-like fermionic dark matter candidate, $\phi$ is the mediator that communicates between the dark sector and the standard model (SM), and H is the SM Higgs doublet.
Therefore $\phi$ should be heavier than $\chi$, i.e., $m_{\phi} > m_{\chi}$. 
$A^{\prime \mu}$ is the gauge boson field corresponding to the $U(1)^{\prime}$ gauge group, with $F_{\mu \nu}^{\prime} = \partial_{[\mu} A_{\nu]}^{\prime}$ being its field strength tensor.
$\chi$, $\phi$ carry the same $U(1)^{\prime}$ charge $Y^{\prime} \neq 0$.
Four-scalar couplings $\kappa_1$ and $\kappa_2$ are inevitable in our model, but for simplicity we just consider the case that these two couplings are small enough and have no influence in our analysis below.
The quantum numbers of the dark sector particles are summarized in Tab.~\ref{Particles}.
\begin{table}
\begin{tabular}{|c|c|c|}
\hline
 & SM $SU(3)_{\text{C}} \times SU(2)_{\text{L}} \times U(1)_Y$ & dark $U(1)^{\prime}$ \\
 \hline
$A_{\mu}^{\prime}$ & (0, 0, 0) & 0 \\
 $\chi$ & (1, 1, 0) & $Y^{\prime}$ \\
 $\phi$ & (1, 1, -1) & $Y^{\prime}$ \\
 \hline
\end{tabular}
\caption{Quantum numbers of the dark sector in the model} \label{Particles}
\end{table}

\begin{figure}[ht]
\includegraphics[width=2in]{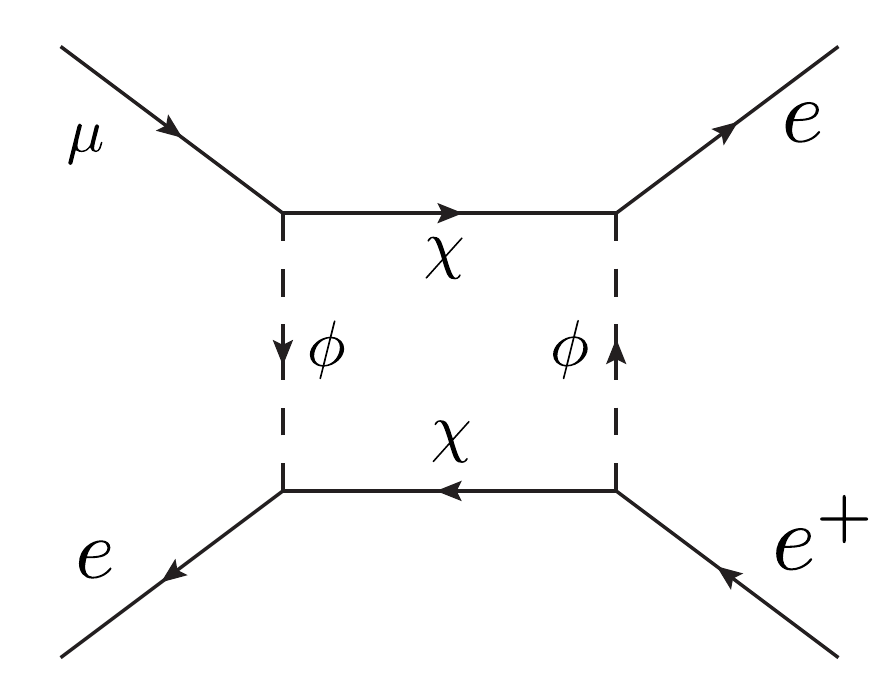}
\includegraphics[width=2in]{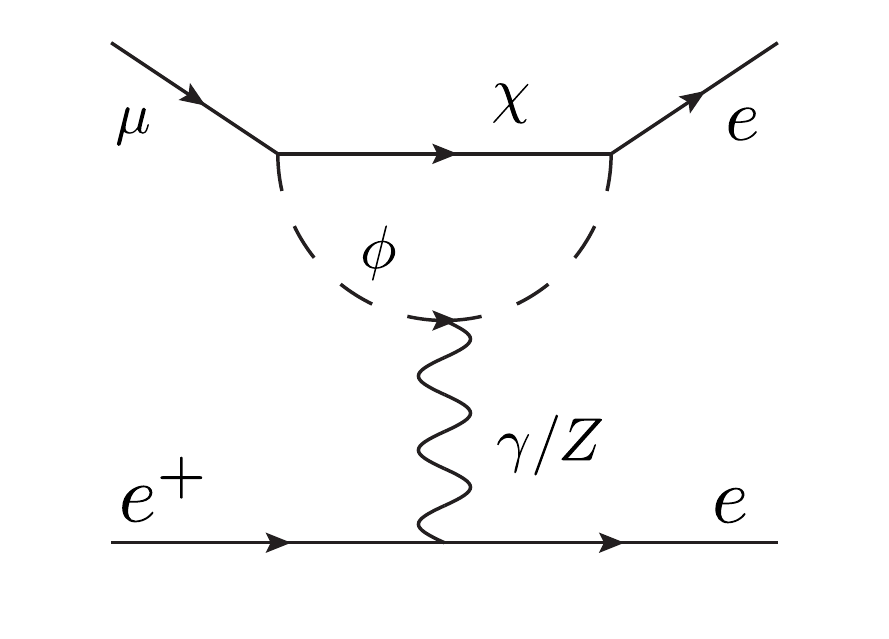}
\caption{Feynman diagrams contributing to the $\mu \rightarrow 3 e$ processes.} \label{mu23e_D}
\end{figure}
In Eq.~(\ref{Lag}) , the coupling $\lambda_i \phi \overline{\chi} P_R e_R^i + \text{h.c.}$ induces the interactions between the SM and the dark sector, where $i$ runs from  $e$, $\mu$, and $\tau$. It should be noted that the non-zero $\lambda$ for three generations will lead to the lepton flavor violating processes at one-loop level, such as $\mu \to 3 e$, as shown in Fig.~\ref{mu23e_D}. The dominant contribution comes from the effective operator
\begin{eqnarray}
\mathcal{H}_{\rm{eff}} \supset \frac{1}{\Lambda^2} \overline{\mu} \gamma^{\mu} P_R e \overline{e} \gamma_{\mu} P_R e,
\end{eqnarray}
with
\begin{eqnarray}
\left| \frac{1}{\Lambda^2} \right| = \frac{\lambda_1^3 \lambda_2}{32 \pi^2} \frac{ (m_{\phi}^2-m_{\chi}^2)(m_{\phi}^2+m_{\chi}^2) + 2 m_{\phi}^2 m_{\chi}^2 \ln \frac{m_{\chi}^2}{m_{\phi}^2} }{2 (m_{\phi}^2 - m_{\chi}^2)^3},
\end{eqnarray}
where we have neglected all terms proportional to $m_{\mu,e}$. Then, we can have the branching ratio of $\mu \rightarrow 3 e$,
\begin{eqnarray}
{\rm Br} (\mu \rightarrow 3 e) \simeq \frac{1}{4 \Lambda^4 G_F^2}.
\end{eqnarray}
A typical result is calculated to be about $\sim 10^{-10}$ in the case that $\lambda_{1,2} \sim 1$ and $m_{\phi} \gtrsim m_{\chi} = 1.5$ TeV, given a correct dark matter relic density. This is about two orders of magnitude larger than the current bound, ${\rm Br} (\mu \rightarrow 3 e) < 10^{-12}$~\cite{PDG}. Therefore, we assume $\lambda_2=\lambda_3=0$ for simplicity in our following calculations.

In Eq.~(\ref{Lag}), we assume a Stueckelberg mass for the dark gauge boson. However, one can introduce a dark Higgs boson to spontaneously break the $U(1)^{\prime}$, which generates a mass term for the gauge boson. If only the nonzero $U(1)^{\prime}$ charge that the dark Higgs boson carries does not  equal $\pm 2 Y^{\prime}$, a global $U(1)^{\prime}_{\text{G}}$ symmetry still remains after the spontaneous breaking of the gauge symmetry. This forbids the Majorana mass terms of the $\chi$ and preserves the stability of the dark matter. The dark sector might also communicate with the SM sector through the kinematic mixing terms $\epsilon B_{\mu \nu} F^{\prime \mu \nu}$, where $B_{\mu \nu} = \partial_{[\mu} B_{\nu]}$ is the gauge field for the hyper-charge. In this paper, we assume this to be small enough to be ignored.

\section{Numerical Results and Discussions}\label{section3}

\begin{figure}[ht]
\includegraphics[height=5.5cm,width=5cm]{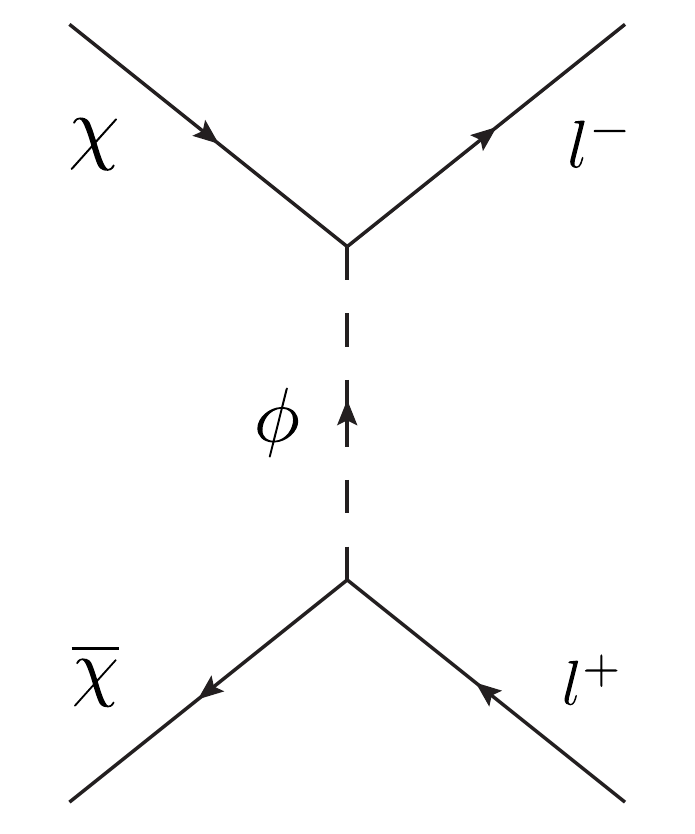}
\vspace{-0.5cm}
\caption{Representative Feynman diagrams of the DM annihilation process.} \label{Annihilation}
\end{figure}

In this paper, we fix the dark matter mass $m_{\chi}$ at1500 GeV, and require that $m_{\phi} > m_{\chi}$. The main annihilation channel is the $t$-channel $\chi \chi \rightarrow e^+ e^-$ shown in the left panel in Fig.~\ref{Annihilation}. We compute the dark matter relic density by micrOMEGAs 4.3.5 \cite{micrOMEGAs} and require our samples to satisfy the DM relic density given by PLANCK data, $\Omega_{\text{DM}} h^2 = 0.1199 \pm 0.0027$ \cite{Planck2013}.

\begin{figure}[ht]
\includegraphics[width=3.2in]{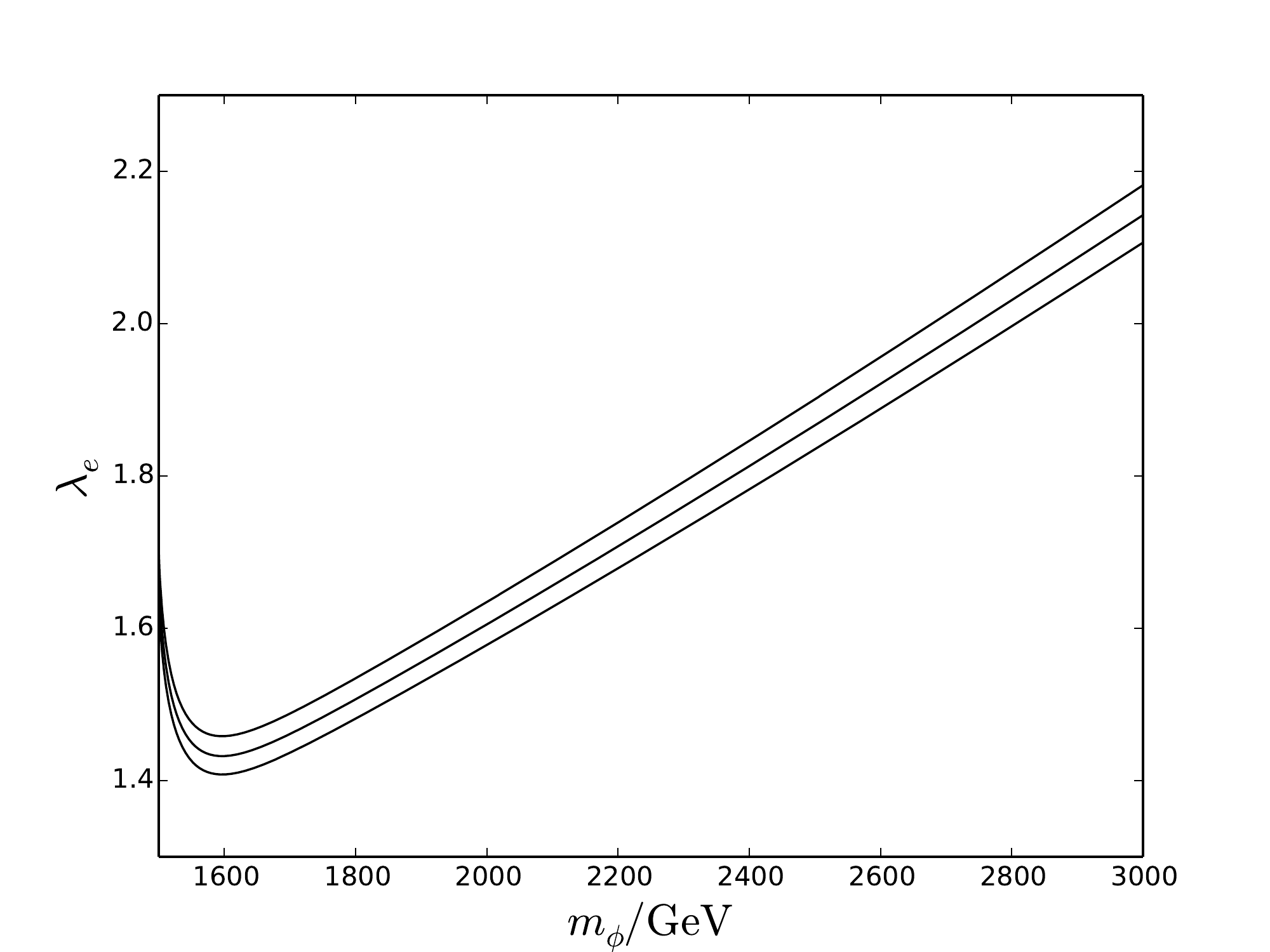}
\includegraphics[width=3.2in]{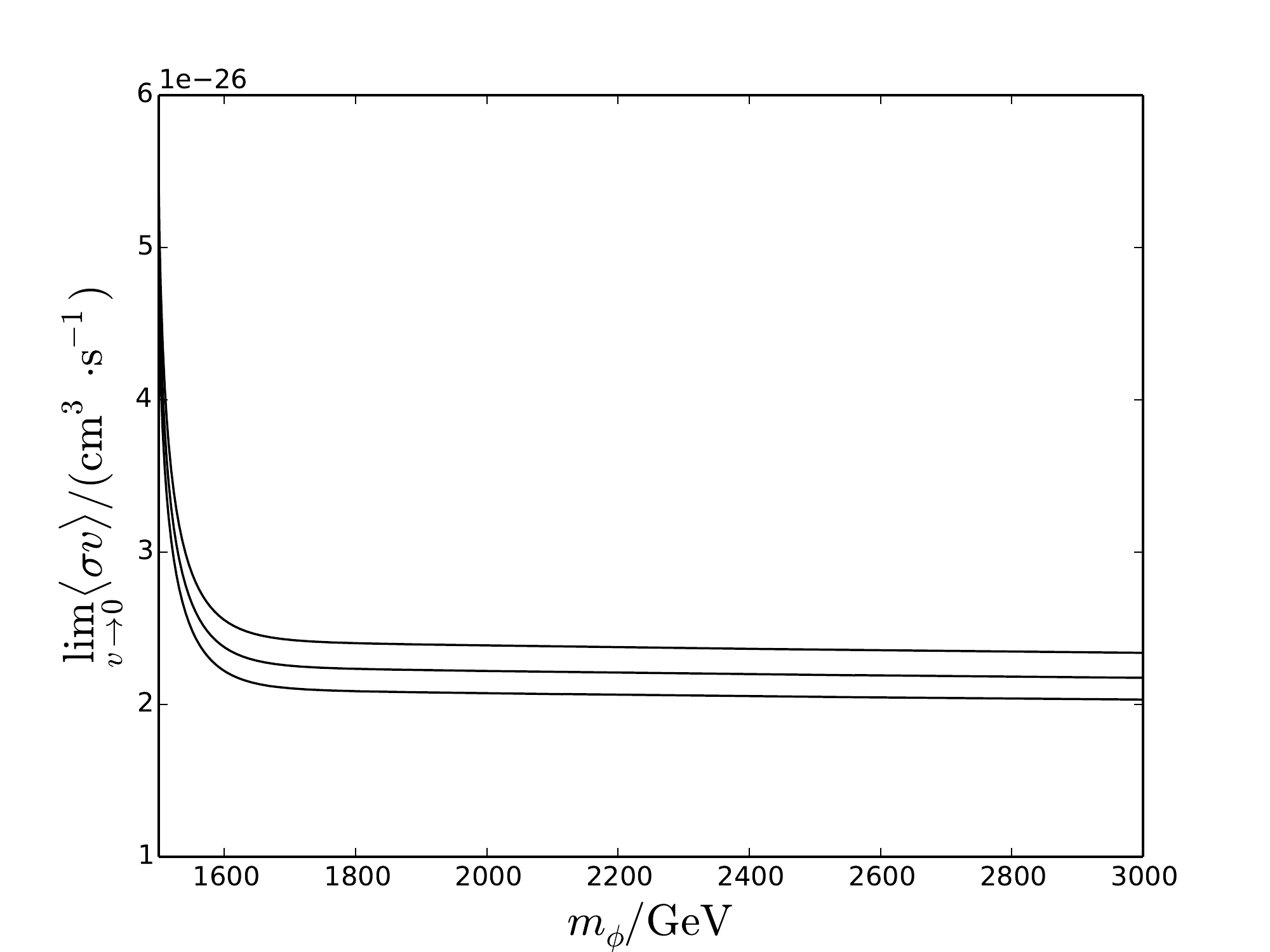}
\caption{Left panel: the contour of DM relic density within $3\sigma$ region of Planck measurement~\cite{Planck2013} on the plane of Yukawa coupling $\lambda_e$ versus mediator mass $m_{\phi}$. Right panel: the dependence of $\langle \sigma v \rangle_{v \to 0}$ on $m_{\phi}$.} \label{Couplings}
\end{figure}
In Fig.~\ref{Couplings}, we plot the contour of DM relic density within $3\sigma$ region of Planck measurement on the plane of Yukawa coupling $\lambda_e$ versus mediator mass $m_\phi$. We can see that the Yukawa coupling $\lambda_e$ has to be larger than 1.4 for the mediator with a mass heavier than 1.5 TeV. Besides, it should be noted that the dip of the curve is caused by the coannihilation effect when the DM mass is close to the mediator mass. We also show the cross section times velocity in the $v \to 0$ limit. We can see that the requirement of $\langle \sigma v \rangle \sim 3 \times 10^{-26} \text{cm}^3/\text{s}$ imposes a strong bound on the mediator mass $m_\phi$, which has to be around 1.5 TeV.

\begin{figure}[ht]
\includegraphics[width=3in]{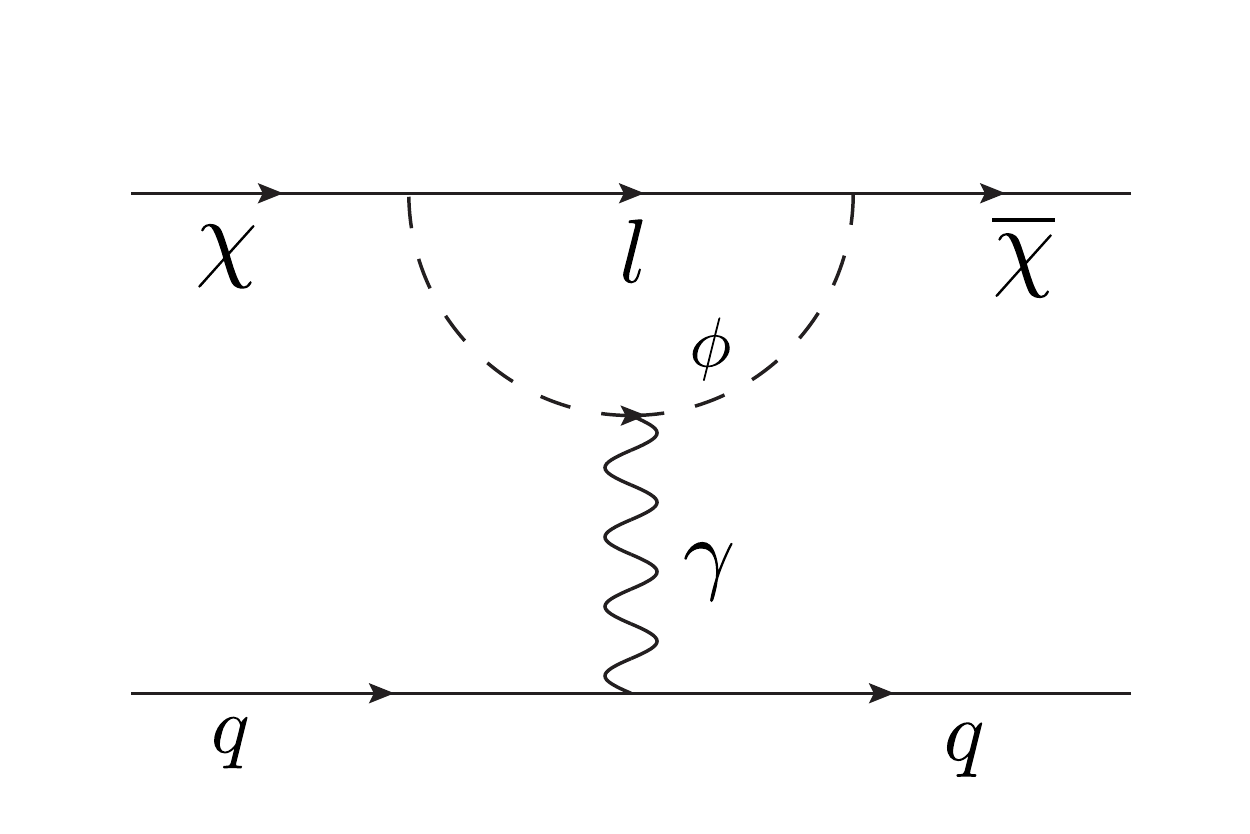}
\caption{Dark matter scattering with the nucleons.} \label{DirectDetections}
\end{figure}

\begin{figure}[ht]
\includegraphics[width=4in]{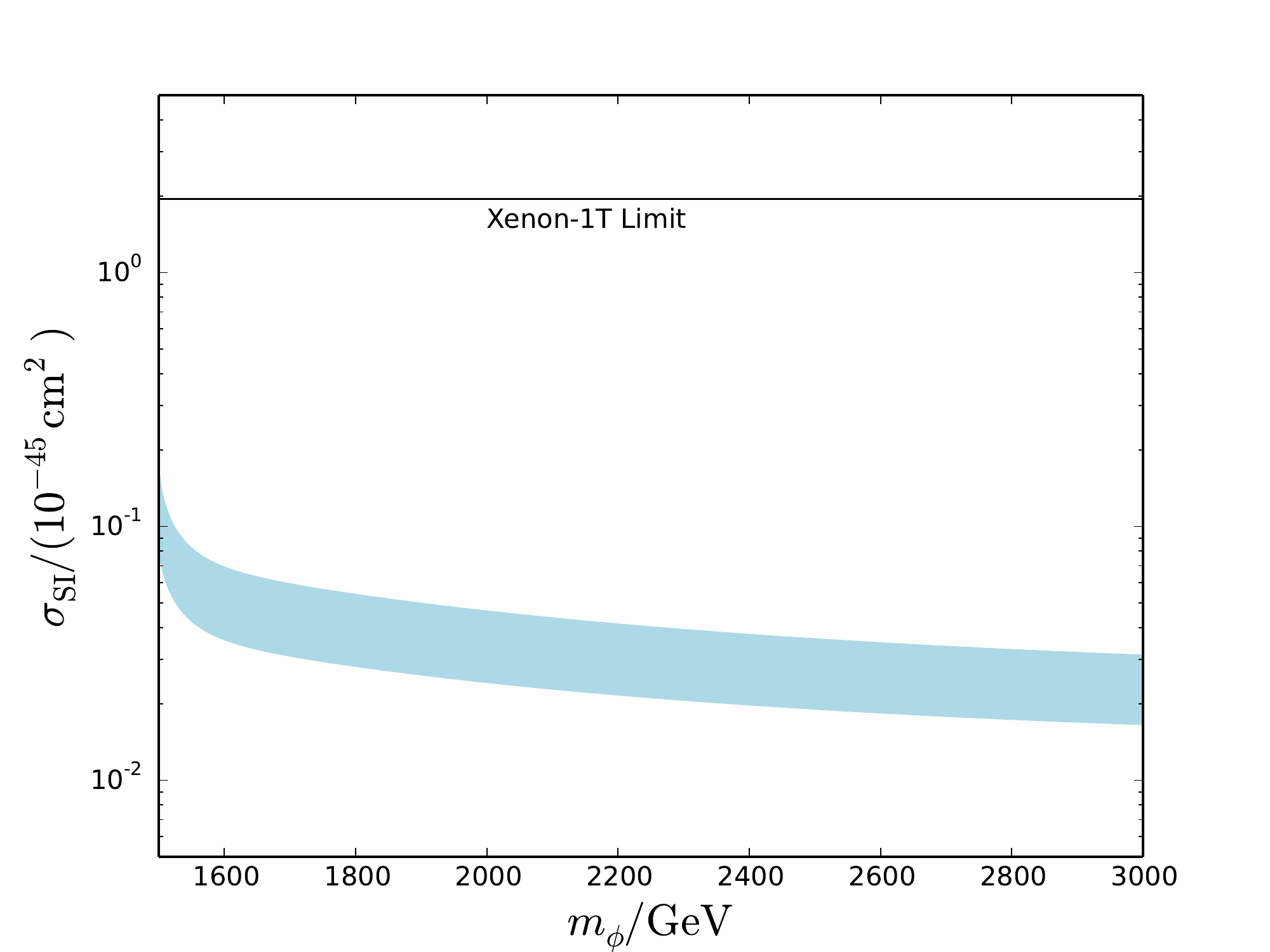}
\caption{The dark matter effective spin-independent cross section with a nucleon, compared with the most stringent Xenon-1T bound.} \label{DDSI}
\end{figure}

Although the lepton-portal DM has no tree-level coupling with the nucleons, it can scatter with the nucleons at one-loop level, as shown in  Fig.~\ref{DirectDetections}. The effective averaged DM-nucleon cross section is given by,
\begin{eqnarray}
\sigma_{\chi N} = c_1^2 e^2 Z^2 \frac{\mu_{\chi N}^2}{A^2 \pi},
\end{eqnarray}
where $\mu_{\chi N}$ is the reduced mass of the DM-nucleon system, $e$ is the electro-magnetic coupling constant, $Z=54$ and $A=129$ are the atomic number and the atomic mass number of a target xenon nucleus. The definition of the $c_1$ is given by~\cite{LeptonPortal}
\begin{eqnarray}
c_1 = \frac{\lambda_e^2 e}{64 \pi^2 m_{\phi}^2} \left[ \frac{1}{2} + \frac{2}{3} \ln \left( \frac{m_e^2}{m_{\phi}^2} \right) \right],
\end{eqnarray}
where $m_e$ originally indicates the mass of an electron. However, as the mass of the electron is below the exchange momentum of the scattering process, $m_e$ should be replaced by the scale $\sim \mu_{\chi T} v$, where $\mu_{\chi T}$ is the reduced mass of the dark matter-xenon target system. Varying $m_e$ from $0.1 \mu_{\chi T} v$ to $\sqrt{2} \mu_{\chi T}$, together with the uncertainty of the dark matter relic density, we give a band of the predicted effective spin-independent DM direct detection cross section with a nucleon in Fig.~\ref{DDSI}. The current most stringent Xenon-1T bound from the Ref.~\cite{xenon-1t} is also plotted there.

The lepton-portal dark matter can also scatter with the electrons. The effective operator can be written as
\begin{eqnarray}
\mathcal{H}_{\text{eff}} \supset \frac{2}{\Lambda^2} (\overline{\chi} P_R e) (\overline{e} P_L \overline{\chi}) = \frac{1}{\Lambda^2} (\overline{\chi} \gamma^{\mu} P_L \chi) (\overline{e} \gamma^{\mu} P_R e).
\end{eqnarray}
Here, Fierz identity has been applied, as well as the approximation $\frac{1}{\Lambda^2} \sim \frac{1}{|m_{\chi}^2+m_e^2+2 m_{\chi} m_e - m_{\phi}^2|}$. The cross section will be significantly amplified if $m_{\phi}$ approaches $m_{\chi}$. However, as long as $m_{\phi} - m_{\chi} > 1$ GeV, and $\frac{1}{\Lambda^2} < \frac{1}{3000} \text{GeV}^{-2}$, then one will have $\sigma_{\chi e} \sim \frac{m_e^2}{\Lambda^4} \sim 10^{-41} \text{ cm}^2$, which is far below the $10^{-34}$-$10^{-32} \text{ cm}^2$ current bound, extracted in Ref.~\cite{Xenon1TLeptonDirect}. On the other hand, for cosmic-ray constraints, it has been shown in~\cite{QiangYuan} that an electron-portal 1.5TeV DM is consistant with Fermi-LAT observation~\cite{Ackermann:2014usa,Abdollahi:2017kyf}. H.E.S.S. gamma-ray limit is also too weak to constrain our scenario~\cite{Abdallah:2016ygi,Profumo:2017obk}.

Next, we discuss the collider phenomenology of our model. Since $\phi$ carries hypercharge, it can be pair-produced through $s$-channel via a photon or a $Z$-boson and then decay into an electron plus a dark matter particle. Due to the relatively large dark matter mass in our scenario, final state $A^{\prime}$ radiation is negligible~\cite{darkshower1,darkshower2}. So this will produce the dilepton with large missing energy signature at the LHC~\cite{ATLASSlepton}. Cross section of $\phi \phi^*$ at 14 TeV LHC is found to be about $2.8 \times 10^{-7} \text{ pb}$, which is much lower than the current sensitivity of LHC searching for two leptons with large missing energy events. Furthermore, we perform the simulations of such process at a 100 TeV hadron collider with  MadGraph5~\cite{MadGraph5}, Pythia6~\cite{PYTHIA6}, Delphes~\cite{Delphes} and FastJet\cite{FastJet}. In order to uncover signals with vastly different mass splitting between the mediator and dark matter, we propose three signal regions, which are called "Loose", "Middle", and "Tight", respectively. Following is our cut flow and categorization criteria:
\begin{itemize}
\item Veto events containing jets or muons with $p_T >$30GeV and $|\eta|<$2.4.
\item Require two electrons with $p_T >$30GeV, $|\eta|<$2.4, and opposite charges.
\item Define two variables $m_{ll}$ and $m_{T2}$ based on the 4-momentum of these two electrons.
Here $m_{ll}$ is the invariant mass of this electron pair, and $m_{T2}$ is defined as:
\begin{eqnarray}
m_{T2} = \mathop{\text{min}}\limits_{\textbf{q}_T} \left[\text{max}\left(m_T(\textbf{p}^{e1}_T,\textbf{q}_T),m_T(\textbf{p}^{e2}_T,\textbf{p}^{miss}_T-\textbf{q}_T)\right)\right],
\end{eqnarray}
with:
\begin{eqnarray}
m_T(\textbf{p}_T,\textbf{q}_T) = \sqrt{2(p_T q_T - \textbf{p}_T \cdot \textbf{q}_T)}.
\end{eqnarray}
\item If an event has $m_{ll}>$100GeV and $m_{T2}>$100GeV, then this event counts as  "Loose";
if this event has $m_{ll}>$300GeV and $m_{T2}>$300GeV, then it counts as "Middle";
if this event has $m_{ll}>$500GeV and $m_{T2}>$500GeV, then it counts as "Tight".
\end{itemize}

The SM background mainly originates from on-shell/off-shell $ZZ$ or $WW$ pair production. The reducible $pp \rightarrow t \overline{t}$ contribution is also taken into account. The expected signal and background event numbers at 20 and 100 ab$^{-1}$ are listed in Tab.~\ref{E_N}. We can see that the significance $S/\sqrt{B}$ can be 2.30$\sigma$ and 5.14$\sigma$ for $m_\phi=2.2$ TeV if the integrated luminosity of a 100 TeV hadron collider can reach 20 ab$^{-1}$ and 100 ab$^{-1}$, respectively.
\begin{table}
\begin{tabular}{|c|c|c|c|c|c|}
\hline Mediator mass & Loose & Middle & Tight & $S/\sqrt{B}$ (20 ab$^{-1}$) &  $S/\sqrt{B}$ (100 ab$^{-1}$) \\\hline
 1600 GeV & 87.8 & 0 & 0 & 0.26 & 0.58 \\\hline
 1800 GeV & 122.4 & 49.2 & 2.6 & 1.23 & 2.75 \\\hline
 2000 GeV & 84.8 & 56.8 & 28.0 & 2.11 & 4.72 \\\hline
 2200 GeV & 59.2 & 45.6 & 30.6 & 2.30 & 5.14 \\\hline
 2400 GeV & 40.4 & 33.6 & 25.6 & 1.92 & 4.29 \\\hline
 SM BKG & 1.15$\times$10$^{5}$ & 1604 & 176 & - & - \\\hline
  \end{tabular}
  \caption{Expected signal and background event number in the signal regions with luminosity 20 ab$^{-1}$ at a 100 TeV hadron collider.}
\label{E_N}
\end{table}

\section{Conclusions}\label{section4}
We investigated the excess of cosmic-ray $e^\pm$ reported by DAMPE in a Hidden Valley Model with the lepton-portal dark matter. The DMs directly annihilate into leptons through $t$-channel. The electron-portal can successfully explain the $e^\pm$ excess and satisfy the DM relic density and direct detection bounds. However, electron+muon/electron+muon+tau-portal is tightly constrained by the lepton flavor violating observables, such as $\mu \to 3 e$. We also analyzed the observability of production process $pp \to \phi\phi^* \to 2e+E^{miss}_T$ in our model and found that it may be probed at a 100 TeV hadron collider.

\begin{acknowledgements}
This work was supported by the National Natural Science Foundation of China (NNSFC) under grants No.~11705093, and the Korea Research Fellowship Program through the National Research Foundation of Korea (NRF) funded by the Ministry of Science and ICT (2017H1D3A1A01014127), and IBS under the project code IBS-R018-D1 (MZ).

\end{acknowledgements}


\begin{thebibliography}{99}

\bibitem{PandaX_SD}
C.~Fu {\it et al.} [PandaX-II Collaboration],
  Phys.\ Rev.\ Lett.\  {\bf 118}, no. 7, 071301 (2017)
  Erratum: [Phys.\ Rev.\ Lett.\  {\bf 120}, no. 4, 049902 (2018)]
  [arXiv:1611.06553 [hep-ex]].

\bibitem{xenon-1t}
  E.~Aprile {\it et al.} [XENON Collaboration],
  Phys.\ Rev.\ Lett.\  {\bf 119}, no. 18, 181301 (2017)
  [arXiv:1705.06655 [astro-ph.CO]].

\bibitem{PAMELA} 
  O.~Adriani {\it et al.} [PAMELA Collaboration],
  Nature {\bf 458}, 607 (2009)
  [arXiv:0810.4995 [astro-ph]].

\bibitem{AMS_02}
  M.~Aguilar {\it et al.} [AMS Collaboration],
  Phys.\ Rev.\ Lett.\  {\bf 113}, 221102 (2014).

\bibitem{Fermi_LAT}
  M.~Ackermann {\it et al.} [Fermi-LAT Collaboration],
  Phys.\ Rev.\ Lett.\  {\bf 115}, no. 23, 231301 (2015)
  [arXiv:1503.02641 [astro-ph.HE]].

\bibitem{CALET}
  O.~Adriani {\it et al.} [CALET Collaboration],
  Phys.\ Rev.\ Lett.\  {\bf 119}, no. 18, 181101 (2017)
  [arXiv:1712.01711 [astro-ph.HE]].

\bibitem{dampemission}
  J.~Chang {\it et al.} [DAMPE Collaboration],
  Astropart.\ Phys.\  {\bf 95}, 6 (2017)
  [arXiv:1706.08453 [astro-ph.IM]].

\bibitem{dampe2017}
  G.~Ambrosi {\it et al.} [DAMPE Collaboration],
  Nature {\bf 552}, 63 (2017)
  [arXiv:1711.10981 [astro-ph.HE]].

\bibitem{QiangYuan}
  Q.~Yuan {\it et al.},
  arXiv:1711.10989 [astro-ph.HE].

\bibitem{Zu:2017dzm} 
  L.~Zu, C.~Zhang, L.~Feng, Q.~Yuan and Y.~Z.~Fan,
  arXiv:1711.11052 [hep-ph].

\bibitem{Gu:2017gle} 
  P.~H.~Gu and X.~G.~He,
  Phys.\ Lett.\ B {\bf 778}, 292 (2018)
  [arXiv:1711.11000 [hep-ph]].

\bibitem{Duan:2017pkq} 
  G.~H.~Duan, L.~Feng, F.~Wang, L.~Wu, J.~M.~Yang and R.~Zheng,
  JHEP {\bf 1802}, 107 (2018)
  [arXiv:1711.11012 [hep-ph]].

\bibitem{Fang:2017tvj} 
  K.~Fang, X.~J.~Bi and P.~F.~Yin,
  Astrophys.\ J.\  {\bf 854}, no. 1, 57 (2018)
  [arXiv:1711.10996 [astro-ph.HE]].

\bibitem{Fan:2017sor} 
  Y.~Z.~Fan, W.~C.~Huang, M.~Spinrath, Y.~L.~S.~Tsai and Q.~Yuan,
  arXiv:1711.10995 [hep-ph].

\bibitem{Athron:2017drj} 
  P.~Athron, C.~Balazs, A.~Fowlie and Y.~Zhang,
  JHEP {\bf 1802}, 121 (2018)
  [arXiv:1711.11376 [hep-ph]].

\bibitem{Chao:2017emq} 
  W.~Chao, H.~K.~Guo, H.~L.~Li and J.~Shu,
  arXiv:1712.00037 [hep-ph].

\bibitem{Huang:2017egk} 
  X.~J.~Huang, Y.~L.~Wu, W.~H.~Zhang and Y.~F.~Zhou,
  arXiv:1712.00005 [astro-ph.HE].

\bibitem{PDG} 
  C.~Patrignani {\it et al.} [Particle Data Group],
  Chin.\ Phys.\ C {\bf 40}, no. 10, 100001 (2016).

\bibitem{micrOMEGAs} 
  D.~Barducci, G.~Belanger, J.~Bernon, F.~Boudjema, J.~Da Silva, S.~Kraml, U.~Laa and A.~Pukhov,
  Comput.\ Phys.\ Commun.\  {\bf 222}, 327 (2018)
  [arXiv:1606.03834 [hep-ph]].

\bibitem{Planck2013}
  P.~A.~R.~Ade {\it et al.} [Planck Collaboration],
  Astron.\ Astrophys.\  {\bf 571}, A16 (2014)
  [arXiv:1303.5076 [astro-ph.CO]].

\bibitem{LeptonPortal}
  Y.~Bai and J.~Berger,
  JHEP {\bf 1408}, 153 (2014)
  [arXiv:1402.6696 [hep-ph]].

\bibitem{Xenon1TLeptonDirect}
  E.~Aprile {\it et al.} [XENON100 Collaboration],
  Science {\bf 349}, no. 6250, 851 (2015)
  [arXiv:1507.07747 [astro-ph.CO]].


\bibitem{Ackermann:2014usa} 
  M.~Ackermann {\it et al.} [Fermi-LAT Collaboration],
  Astrophys.\ J.\  {\bf 799}, 86 (2015)
  [arXiv:1410.3696 [astro-ph.HE]].

\bibitem{Abdollahi:2017kyf} 
  S.~Abdollahi {\it et al.} [Fermi-LAT Collaboration],
  Phys.\ Rev.\ Lett.\  {\bf 118}, no. 9, 091103 (2017)
  [arXiv:1703.01073 [astro-ph.HE]].

\bibitem{Abdallah:2016ygi} 
  H.~Abdallah {\it et al.} [H.E.S.S. Collaboration],
  Phys.\ Rev.\ Lett.\  {\bf 117}, no. 11, 111301 (2016)
  [arXiv:1607.08142 [astro-ph.HE]].


\bibitem{Profumo:2017obk} 
  S.~Profumo, F.~S.~Queiroz, J.~Silk and C.~Siqueira,
  JCAP {\bf 1803}, no. 03, 010 (2018)
  [arXiv:1711.03133 [hep-ph]].

\bibitem{darkshower1} 
  M.~Buschmann, J.~Kopp, J.~Liu and P.~A.~N.~Machado,
  JHEP {\bf 1507}, 045 (2015)
  [arXiv:1505.07459 [hep-ph]].

\bibitem{darkshower2} 
  M.~Zhang, M.~Kim, H.~S.~Lee and M.~Park,
  arXiv:1612.02850 [hep-ph].

\bibitem{ATLASSlepton} 
  The ATLAS collaboration [ATLAS Collaboration],
  ATLAS-CONF-2017-039.

\bibitem{MadGraph5} 
  J.~Alwall {\it et al.},
  JHEP {\bf 1407}, 079 (2014)
  [arXiv:1405.0301 [hep-ph]].

\bibitem{PYTHIA6} 
  T.~Sjostrand, S.~Mrenna and P.~Z.~Skands,
  JHEP {\bf 0605}, 026 (2006)
  [hep-ph/0603175].

\bibitem{Delphes} 
  J.~de Favereau {\it et al.} [DELPHES 3 Collaboration],
  JHEP {\bf 1402}, 057 (2014)
  [arXiv:1307.6346 [hep-ex]].

\bibitem{FastJet} 
  M.~Cacciari, G.~P.~Salam and G.~Soyez,
  Eur.\ Phys.\ J.\ C {\bf 72}, 1896 (2012)
  [arXiv:1111.6097 [hep-ph]].

\end{thebibliography}
\end{document}